\documentclass[fleqn,usenatbib]{mnras}

\usepackage{newtxtext,newtxmath}
\usepackage[T1]{fontenc}

\DeclareRobustCommand{\VAN}[3]{#2}
\let\VANthebibliography\thebibliography
\def\thebibliography{\DeclareRobustCommand{\VAN}[3]{##3}\VANthebibliography}

\usepackage{xcolor} % For comments with different colours

\usepackage{graphicx}	% Including figure files
\usepackage{amsmath}	% Advanced maths commands
\newcommand{\targ}{J0240+1952}
\newcommand{\target}{LAMOST J024048.51+195226.9}

\title[The rapid spin of \target]{Found: a rapidly spinning white dwarf in \target}

\author[Pelisoli et al.]{
Ingrid Pelisoli,$^{1}$\thanks{E-mail: ingrid.pelisoli@warwick.ac.uk}
T.~R.\ Marsh,$^{1}$
V.~S.\ Dhillon,$^{2,3}$
E. Breedt,$^{4}$
A.~J. Brown,$^{2}$
M.~J. Dyer,$^{2}$
M.~J. Green,$^{5}$\newauthor
P. Kerry,$^{2}$
S.~P. Littlefair,$^{2}$
S.~G. Parsons,$^{2}$
D.~I. Sahman,$^{2}$
J.~F. Wild$^{2}$
\\
$^{1}$Department of Physics, University of Warwick, Gibbet Hill Road, Coventry, CV4 7AL, UK\\
$^{2}$Department of Physics and Astronomy, Hicks Building, The University of Sheffield, Sheffield, S3 7RH, UK\\
$^{3}$Instituto de Astrof\'{i}sica de Canarias, E-38205 La Laguna, Tenerife, Spain\\
$^{4}$Institute of Astronomy, University of Cambridge, Madingley Road, Cambridge CB3 0HA, UK\\
$^{5}$Department of Astrophysics, School of Physics and Astronomy, Tel Aviv University, Tel Aviv 6997801, Israel
}

\date{Accepted XXX. Received YYY; in original form ZZZ}

\pubyear{2015}

\begin{document}
\label{firstpage}
\pagerange{\pageref{firstpage}--\pageref{lastpage}}
\maketitle

\begin{abstract}
We present optical photometry of the cataclysmic variable \target\ taken with the high-speed, five-band CCD camera HiPERCAM on the 10.4\,m Gran Telescopio Canarias (GTC). We detect pulsations originating from the spin of its white dwarf, finding a spin period of 24.9328(38)\,s. The pulse amplitude is of the order of 0.2 per cent in the $g$-band, below the detection limits of previous searches. This detection establishes \target\ as only the second white dwarf magnetic propeller system, a twin of its long-known predecessor, AE~Aquarii. At 24.93\,s, the white dwarf in \target\ has the shortest known spin period of any cataclysmic variable star. The white dwarf must have a mass of at least $0.7$\,$\mathrm{M}_{\sun}$ to sustain so short a period. The observed faintest $u$-band magnitude sets an upper limit on the  white dwarf's temperature of $\sim 25\,000$\,K. The pulsation amplitudes measured in the five HiPERCAM filters are consistent with an accretion spot of $\sim 30\,000\,$K covering $\sim 2$ per cent of the  white dwarf's visible area, although spots that are hot and smaller, or cooler and larger cannot be ruled out. 
\end{abstract}

\begin{keywords}
binaries: general -- stars: cataclysmic variables -- binaries: close
\end{keywords}

%%%%%%%%%%%%%%%%%%%%%%%%%%%%%%%%%%%%%%%%%%%%%%%%%%

%%%%%%%%%%%%%%%%% BODY OF PAPER %%%%%%%%%%%%%%%%%%

\section{Introduction}

% Paragraph on magnetic CVs and their importance
Magnetic fields play a central role in the evolution and observational properties of close interacting binaries, in particular the cataclysmic variables \citep[CVs; e.g.][]{Schreiber2021}. In these systems, a white dwarf accretes mass from a late-type main-sequence star via Roche lobe overflow. If the white dwarf is magnetic, the magnetic field can regulate the geometry and even the rate of accretion, and causes distinct observational properties, which characterise the main types of CVs. For low magnetic fields ($B \lesssim  1\,$MG), accretion takes place in a disc extending down to the white dwarf's equator. At the opposite extreme, for magnetic fields $B \gtrsim 10\,$MG, the accretion disc can be entirely disrupted, with accretion occurring along magnetic field lines and the white dwarf's spin is locked to its binary companion in systems known as "polars" \citep[for a thorough review, see][]{Cropper1990}. This leaves the most complex case, the intermediate polars (IPs), which have magnetic fields strong enough to disrupt the inner accretion disc, but insufficient to lock the white dwarf's spin to the orbit. Accretion towards the magnetic poles breaks azimuthal symmetry leading to photometric modulation at the white dwarf's spin period \citep{Patterson1994}.

% AE Aqr stands out
The system AE~Aquarii, which was one of the first to be recognised as a CV and which was of importance in the development of our picture of CVs \citep{Joy1943,Crawford1956}, has long stood out from the crowd of magnetic systems, including its closest cousins, the IPs. AE~Aqr's optical and ultraviolet light curves show irregular flaring of up to a magnitude on time scales of minutes \citep[e.g.][]{Chincarini1981, Welsh1998, Watson2006}. The flares are larger and different in character from the stochastic flickering exhibited by many CVs \citep{Bruch1991}. The flares are also seen spectroscopically in increased fluxes and widths of the Balmer lines \citep{Reinsch1994}. Rounding off the list of notable properties, AE~Aqr was one of the first CVs to be detected at radio frequencies \citep{Bookbinder1987}, and is one of the most luminous of all CVs in terms of radio power \citep{Pretorius2021}. The radio flux is thought to be generated by synchrotron emission from expanding clouds of plasma \citep{Meintjes2003}.

A crucial breakthrough in our understanding of AE~Aqr came with the discovery of coherent pulsations on a period of $33.08$\,s \citep{Patterson1979}, interpreted as the spin of a magnetic white dwarf. It was later discovered that the white dwarf's spin is slowing on a timescale of $P/\dot{P} \sim 10^7\,$yr, meaning that more than enough rotational energy is lost from the white dwarf's spin to power the system \citep{deJager1994,Mauche2006}. Combining this discovery with AE~Aqr's other properties, led to a model for AE~Aqr as a white dwarf magnetic propeller in which the majority of the mass transferred from the secondary star is flung out of the system as it interacts with the white dwarf's magnetosphere, resulting in the flaring and synchrotron emission previously described \citep{Eracleous1996,Wynn1997,Pearson2003,Meintjes2003}. The energy and angular momentum for this process come from the white dwarf's spin.

For over seven decades since its identification as a cataclysmic variable star, AE~Aqr has stood alone amongst the thousands of known systems for this remarkable suite of properties. This might finally have changed with the discovery of a possible twin of AE~Aqr, the star \target\ (henceforth \targ) \citep{Thorstensen2020}. \targ\ was identified as a CV by \citet{Hou2020} from an automated search in spectra obtained by the Large Sky Area Multi-Object Fibre Spectroscopic Telescope \citep[LAMOST,][]{Cui2012}. It had been previously identified as photometrically variable by \citet{Drake2014} using data from the Catalina Real-Time Transient Survey \citep[CRTS,][]{Drake2009}, who found a period of 0.3056840\,d. Follow-up optical spectroscopy and photometry was obtained by \citet{Thorstensen2020}, who noticed striking similarities between \targ\ and AE~Aqr, in particular in its irregular flaring and spectral properties. The spectra of \targ\ are dominated by strong Balmer lines in emission, with the continuum showing wide absorption bands attributed to an M1.5($\pm$1) dwarf companion. \citet{Thorstensen2020} measured radial velocities from the absorption features, and showed that the period identified by \citet{Drake2014} corresponded to the orbital period of the binary. This period was refined by \citet{Littlefield2020} to a precise value of 0.3056849(5)\,d. They also identified shallow eclipses of the white dwarf by the secondary in CRTS data, characterised by a lack of flaring activity. Two subsequent studies have added additional weight to the connection between \targ\ and AE~Aqr. \cite{Pretorius2021} detected radio emission, with a radio luminosity close to the highest measured of any CV ($2.7\pm0.3 \times 10^{17}$\,erg\,s$^{-1}$\,Hz$^{-1}$ measured in the L-band, which is centred at 1284\,MHz), higher even than AE~Aqr, while  \citet{Garnavich2021} found that the Balmer emission lines broadened up to $\pm 3000$\,km/s during flares, and showed that a P~Cygni-like component in H$\alpha$, also seen by \citet{Thorstensen2020}, was consistent with \citet{Wynn1997}'s magnetic propeller model.

While the weight of evidence points strongly towards an AE~Aqr-like magnetic propeller model for \targ, no signs of a fast-spinning white dwarf analogous to the pulsations displayed by AE~Aqr have been reported. \citet{Thorstensen2020} found no evidence of pulsations in 23.3\,s-cadence photometry. \citet{Pretorius2021} searched for optical pulsations down to a period of 10\,s, but found no signals above their detection threshold of $1$ per cent.  Finally, and most stringently of all, \citet{Garnavich2021} presented $g$ and $i$-band photometry taken with the Hale 5\,m telescope and Caltech HIgh-speed Multi-color camERA (CHIMERA) which had the cadence and the signal-to-noise to detect spin periods with amplitudes as low as 4\,mmag ($0.43$ per cent) in the $g$-band over a period range of 6.3 to 85\,s, but again found nothing of significance.

In this paper we present new high-speed and high signal-to-noise optical photometry of \targ, revealing the spin period of its white dwarf for the first time, thereby securing its place as the second white dwarf magnetic propeller system to have been discovered.

\section{Observations \& data reduction}

\targ\ was observed with the high-speed CCD camera HiPERCAM \citep{Dhillon2021}, mounted on the $10.4\,$m Gran Telescopio Canarias (GTC), during the night of 2021 August 7. HiPERCAM uses four dichroic beam splitters to allow simultaneous observations in five different filters, $u_s$, $g_s$, $r_s$, $i_s$, and $z_s$. These filters, which were specifically designed for HiPERCAM, match the cut-on/off wavelengths of the Sloan Digital Sky Survey (SDSS) filters, but have a higher throughput. We used 3x3 CCD binning, resulting in a scale of 0.24~arcsec/pixel. The exposure time was set to 2.36\,s in $g_s$, $r_s$, $i_s$, and $z_s$, and 4.72\,s in $u_s$. We used the slow read out mode, resulting in a dead time of 7.8\,ms between exposures so that we were collecting light for 99.7 per cent of the time. The data were taken in a continuous run of $2.8\,$hours in duration, covering roughly one third of  \targ's orbit. Conditions were clear with seeing of 0.6--1.0~arcseconds.

\begin{figure}
	\includegraphics[width=\columnwidth]{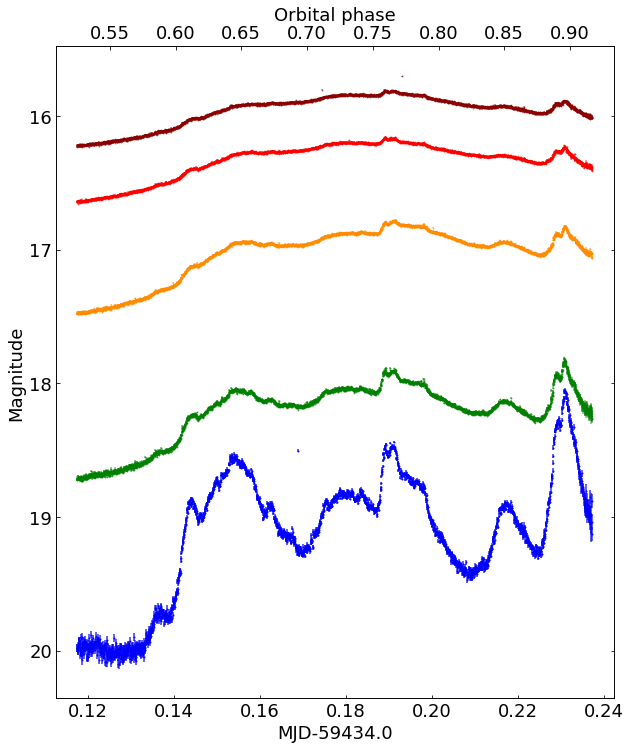}
    \caption{Differential light curves obtained with HiPERCAM simultaneously in the $u_s$, $g_s$, $r_s$, $i_s$, and $z_s$ bands (from bottom to top), showing intense flaring activity. The orbital phase shown on the top x-axis was calculated using the ephemeris of \citet{Garnavich2021}.}
    \label{fig:lightcurve}
\end{figure}

The data were reduced with the dedicated HiPERCAM data reduction pipeline\footnote{https://github.com/HiPERCAM/hipercam}. We first performed bias subtraction, and flat field correction using skyflats taken during twilight. Fringe correction was performed for the $z_s$-band using archive fringe maps. Next we carried out differential aperture photometry using a variable aperture size, set to scale with the seeing measured from a point-spread function (PSF) fit. The bright star Gaia EDR3~84846258594393472 ($u = 17.27, g = 15.69, r = 15.06, i = 14.83, z = 14.79$) was used as comparison, and to obtain magnitude values calibrated to the SDSS system. The resulting differential light curves for the five filters are shown in Fig.~\ref{fig:lightcurve}. We show the $x$-axis both in terms of time and orbital phase, using the ephemeris derived by \citet{Garnavich2021}. After a quiet beginning, dominated by a rise in flux in the redder filters which is probably largely the result of tidal deformation of the secondary star, multiple flares can be seen in the light curves, increasing in relative strength from red to blue. The $u_s$-band light curve is particularly spectacular, with a final flare that rises by over a magnitude above an already flare-enhanced baseline.

\section{The spin of the white dwarf}

\begin{figure}
	\includegraphics[width=\columnwidth]{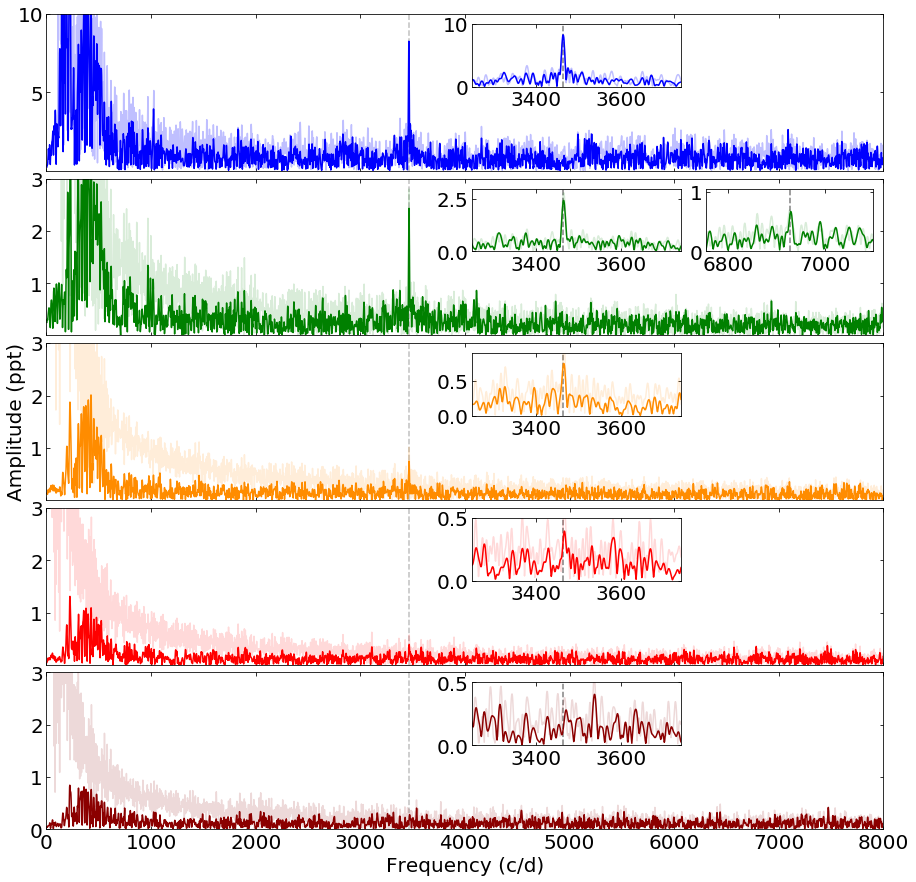}
    \caption{Fourier transform for the five HiPERCAM bands ($u_s$, $g_s$, $r_s$, $i_s$, and $z_s$ from top to bottom) around the identified peak. The lighter lines and darker lines show the amplitude before and after a spline fit to remove flares and orbital variability is subtracted. The amplitude is shown in parts per thousand (ppt), and the frequency in cycles per day (c/d). The insets in the upper right show a zoom around the detected peak. For $g_s$, there is a hint of the first harmonic, shown in a second inset.}
    \label{fig:ft}
\end{figure}

In order to search for the spin of the white dwarf, we performed a Fourier transform (FT) of the light curves up to the Nyquist frequency. The results are shown in Fig.~\ref{fig:ft}, focusing on the region where a peak was identified. The lighter lines show the FT using directly the data shown in Fig.~\ref{fig:lightcurve}, whereas for the darker lines we have subtracted a spline fit to remove orbital and flaring signatures prior to performing the FT. In both cases, there is a signal at $\sim$3500\,cycles per day (c/d, $\sim$25\,s), clearly detected in the $u_s$, $g_s$, and $r_s$ bands. The signal is also marginally visible in the $i_s$-band, but it is not detected  in the $z_s$-band. Such a short period, coherent signal, with an amplitude that increases towards the bluer bands where the white dwarf grows in significance against the light of its companion, is the exact signature expected of a rapidly-spinning white dwarf.

\begin{figure}
	\includegraphics[width=\columnwidth]{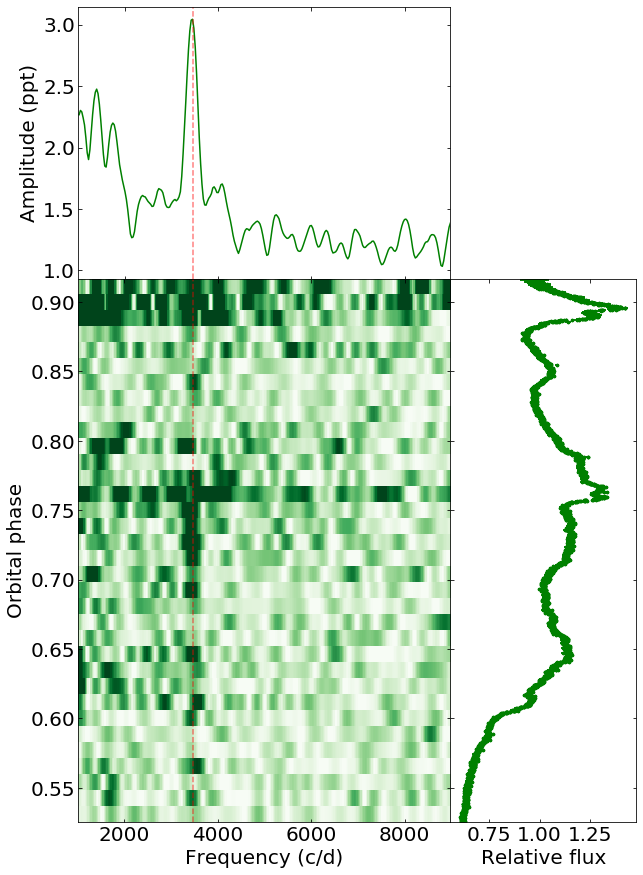}
    \caption{The bottom left panel shows the Fourier transform of independent chunks of data in the $g_s$-band spanning 5 minutes as a function of the orbital phase. The colour scale represents the amplitude in ppt. This amplitude is relative to the overall mean level, not the local mean, and thus is an accurate representation of the variation in absolute strength of the pulsations. The top panel shows the average amplitude over all data chunks. The bottom right panel shows the light curve also as a function of orbital phase. The vertical dashed line in the left panels marks our derived spin period.}
    \label{fig:ft_orb}
\end{figure}

The signal can at times be detected within intervals as short as 5 minutes. This is illustrated in Fig.~\ref{fig:ft_orb}, where we have split the spline-subtracted $g_s$ light curve into independent 5-minute intervals and performed the FT for each of these. A trail can be seen at $\sim$3500\,c/d, with an amplitude typically 3--4 standard deviations above the average amplitude level. Both this ``running'' FT and the light curve are shown as a function of orbital phase. The signal can increase in strength during times of flares (e.g. near phase 0.75), as previously seen for AE~Aqr \citep[see e.g. fig. 8 of][]{Skidmore2003}, though not in a consistent manner, and sometimes the detection of the signal can instead be prevented by the occurrence of strong flares, which can raise the noise level to similar amplitudes to the signal itself, as occurs for the strong flare towards the end of our run.

Once we established the presence of the spin signal in the data, we proceeded to estimate a precise spin period by fitting a cosine to the spline subtracted data. Even though the signal has a stronger amplitude in $u_s$, the $u_s$ data are typically noisier (see the average FT amplitudes in Fig.~\ref{fig:ft}) and more affected by flares, which led to a less precise period than obtained from the $g_s$-band. Therefore, we used the $g_s$-band to derive a spin ephemeris. We fitted the $g_s$-band data using a cosine with the amplitude, time of maximum $T_0$, and period left as free parameters. Observing times were corrected to Barycentric Julian Date (BJD) in the Barycentric Dynamical Time (TDB) reference system. We performed the fit using a least-squares method with the peak of the FT to set initial values for period and amplitude. The times were offset by the value at the middle of our run prior to the fit, and we used zero as the initial guess for $T_0$. Uncertainties were derived via bootstrapping, i.e. the data were re-sampled allowing for repetitions and then refitted, and the standard deviation of each parameter after a thousand fits was taken as the uncertainty. We obtained
\begin{eqnarray}
T_{max} = 2459434.6780256(48) + 0.000288574(44)E
\end{eqnarray}
where $T_{max}$ is the time of maximum expressed in BJD(TDB), and $E$ is the integer cycle count.

The amplitude in the $g_s$-band is found to be $2.30\pm0.20$\,ppt (parts-per-thousand), or $0.230\pm0.020$ per cent, below the detection limits of \citet{Pretorius2021} and \citet{Garnavich2021}. In order to determine the amplitudes in the other bands, we fitted the light curves with the period fixed to the value determined from the $g_s$ data. We used the $T_0$ value determined from $g_s$ as the initial guess, but allowed it to vary to accommodate possible phase shifts between the data taken with different filters. The FT peak amplitude was used as initial guess for the amplitude. As before, uncertainties were determined via bootstrapping. We found the amplitude to be $7.2\pm0.9$\,ppt, $0.66\pm0.12$\,ppt, $0.34\pm0.10$\,ppt, and $0.11\pm0.13$\,ppt in the $u_s$, $r_s$, $i_s$, and $z_s$ bands, respectively. The value of $T_0$ for all bands was consistent with the value derived from the $g_s$-band data.

\section{Discussion}

The measured spin period of 24.93\,s provides the missing evidence that confirms \targ\ as a twin of AE~Aqr, and distinguishes it as the fastest spinning white dwarf yet found in a CV. Importantly, the detection of the spin also confirms that the compact object is indeed a WD and the system is a CV. Without this detection, the observed properties were not dissimilar from transitional millisecond pulsars harbouring neutron stars \citep[e.g.][]{Kennedy2020}.

Only three other systems with spin periods below 40\,s are known, one of them being AE~Aqr itself. The fastest known before the discovery of \targ's spin was CTCV~J2056-3014, which shows a spin period of 29.61\,s \citep{LopesOliveira2020}, whereas the white dwarf in V1460~Her was recently found to show a spin period of 38.75\,s \citep{Ashley2020}. The magnetic fields of the white dwarfs in these latter two, however, are apparently not strong enough to power a propeller. There is a possible faster spinning white dwarf than \targ\ in the wind accreting X-ray binary HD49798 \citep{Israel1997,Mereghetti2011}, though the nature of the compact object as a white dwarf or neutron star is still under discussion \citep{Mereghetti2016, Brooks2017}.

\begin{figure}
	\includegraphics[width=\columnwidth]{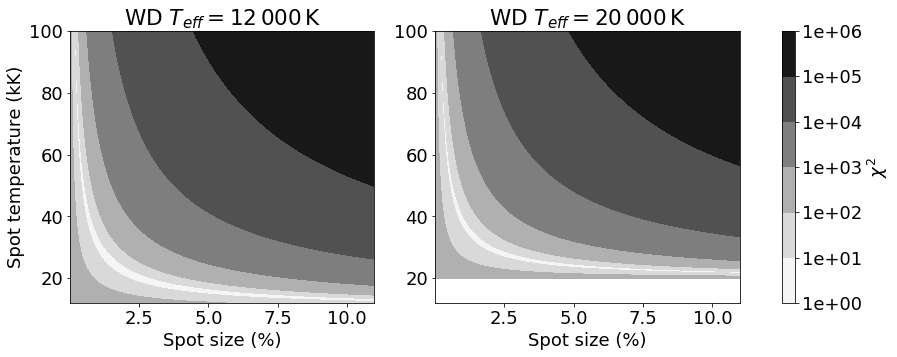}
    \caption{Values of $\chi^2$ as a function of spot temperature and size for two white dwarf temperatures. The $\chi^2$ behaviour suggests that spots larger than 10 per cent, or hotter than 80\,000, are unlikely. Spots cooler than the white dwarf would require an unphysical negative area fraction, and are therefore not considered.}
    \label{fig:chimap}
\end{figure}

\begin{figure}
	\includegraphics[width=\columnwidth]{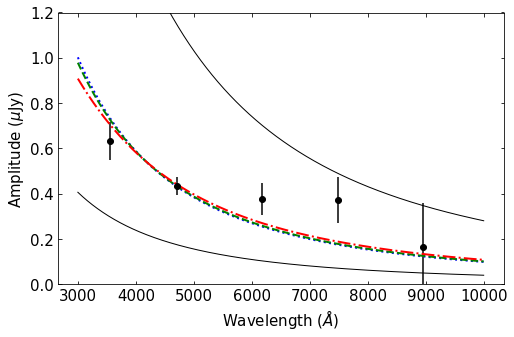}
    \caption{Measured amplitudes, in $\mu$Jy, compared to the difference between the flux of a spot and the flux of the underlying white dwarf. Both are described by blackbodies with a typical white dwarf radius of 0.01\,$\mathrm{R}_{\sun}$ at the distance of \targ. We assume a temperature of 20\,000\,K for the white dwarf, and different temperatures for the spot. The blue dotted line shows a 90\,000\,K blackbody with 0.25 per cent of the white dwarf's projected area, the green dashed line shows 60\,000\,K and 0.44 per cent of the area, and the red dot-dashed line shows 30\,000\,K and 1.9 per cent. The black solid lines illustrate the dramatic change in $\chi^2$ for larger or smaller spots in this temperature range. The upper line is a 30\,000\,K spot with a size of 5 per cent, whereas the lower line shows a 90\,000\,K spot with a size of only 0.1 per cent.}
    \label{fig:bb_amp}
\end{figure}

The measured amplitudes of the pulsations can in principle place constraints on the temperature and size of the spot induced by the small amount of material accreted on the white dwarf's surface. Fig.~\ref{fig:chimap} shows the $\chi^2$ obtained for spot models with a range of temperatures and sizes for two different WD temperatures, 12\,000\,K \citep[near the lowest value observed for CVs, e.g.][]{Pala2020} and 20\,000\,K (consistent with an upper limit derived below).  We assume a radius of 0.01\,$\mathrm{R}_{\sun}$ for the white dwarf, typical for WDs with masses in the range observed for CVs \citep[see e.g. fig. 9 in][]{Parsons2017}, and place it at the distance of \targ, which is $618\pm27$\,pc according to its {\it Gaia} EDR3 parallax \citep{edr3}. It can be seen that these data are insufficient to fully constrain the temperature and size of the spot, which are strongly correlated. The observed amplitudes can be explained by spots slightly hotter than the WD with a few per cent of the its projected area, or by much smaller spots with hotter temperatures. However, the $\chi^2$ values increase dramatically for spots larger than $\approx 10$ per cent, or hotter than $\approx 80\,000$\,K. As a comparison, the spot in AE~Aqr is found to be well described by temperatures in the range $24\,000 - 26\,000$\,K and sizes of a few per cent of the WD area \citep{Eracleous1994}. Fig.~\ref{fig:bb_amp} illustrates some possible models for \targ's spot, showing the amplitudes in $\mu$Jy compared to the difference between the flux of a blackbody describing the spot (with temperatures of 90\,000, 60\,000, and 30\,000\,K) and the flux of an underlying 20\,000\,K white dwarf. The flux was normalised to obtain the minimum $\chi^2$, with the normalisation factor being equivalent to the half the fraction of the white dwarf's projected area that is covered by the spot (our model explains the amplitude itself, whereas the full variability is twice that). For these three temperatures, the spot varies from 1.9 to 0.25 per cent of the white dwarf's area. We also show two models close to the limiting values for the size and temperature of the spot according to the $\chi^2$ behaviour seen in Fig.~\ref{fig:chimap}, to illustrate how these extremes lead to a poor fit.

The temperature of the white dwarf itself can also be constrained from our data. Fig.~\ref{fig:lightcurve} shows that the $u$-band magnitude can be as faint as 20~mag. This implies an upper limit on the white dwarf temperature, above which the white dwarf alone would be brighter than the maximum observed magnitude. Comparing the observed apparent $u$ magnitude with absolute magnitudes for white dwarf models\footnote{https://www.astro.umontreal.ca/~bergeron/CoolingModels/} \citep{Holberg2006,Tremblay2011,Bedard2020}, we find the requirement of $T_\textrm{eff} \lesssim 25\,000$\,K at \targ's distance. This limit depends on the mass of the white dwarf; we have assumed $M\simeq 0.8\,\mathrm{M}_{\sun}$, which is near the mean mass of white dwarfs in CVs \citep{Pala2020}. From the derived spin period, a lower limit on the mass can be placed, below which the white dwarf would be unstable if rotating at such a high rate. According to \citet{Chanmugam1987}, the minimum mass $M$ of a uniformly rotating spheroid with mean radius $R$ and spin period $P$ is such that 
\begin{eqnarray}
\frac{M}{R^3} \geq \frac{117.5}{GP^2},
\end{eqnarray}
where $G$ is the gravitational constant. For our maximum temperature, this implies a minimum mass of $0.7\,\mathrm{M}_{\sun}$, in agreement with the fact that white dwarfs in CVs are found to be typically more massive than the observed field white dwarf mean mass of $0.6\,\mathrm{M}_{\sun}$ \citep[e.g.][]{McCleery2020}, but also suggesting that the white dwarf might not be far from its maximum spin rate.

\section{Summary \& Conclusions}

We have detected a coherent periodic signal with period $P = 24.93\,$s in optical light curves of the accreting white dwarf binary, \mbox{\target}. The signal is consistent with the rapidly spinning white dwarf predicted for the system and firmly establishes \target\ as a twin of the hitherto unique system, AE~Aqr. We have used these data to place constraints on the WD's mass and temperature, as well as on the properties of the accretion-induced spot. Like AE~Aqr, \target\ shows properties fully consistent with a magnetic propeller state in which most mass transferred from its secondary star is ejected from the system. Future measurement of the white dwarf's rate of change of spin period is of very high interest, given the known rapid spin-down of its twin AE~Aqr.

\section*{Acknowledgements}

We thank the anonymous referee for insightful comments that helped improve this manuscript. IP and TRM acknowledge support from the UK's Science and Technology Facilities Council (STFC), grant ST/T000406/1, and from a Leverhulme Research Fellowship. SGP acknowledges the support of a STFC Ernest Rutherford Fellowship. This work is based on observations made with the Gran Telescopio Canarias (GTC), installed at the Spanish Observatorio del Roque de los Muchachos of the Instituto de Astrofísica de Canarias, in the island of La Palma. The design and construction of HiPERCAM was funded by the European Research Council under the European Union’s Seventh Framework Programme (FP/2007-2013) under ERC-2013-ADG Grant Agreement no. 340040 (HiPERCAM). VSD and HiPERCAM operations are supported by STFC grant ST/V000853/1. We would like to thank the staff of the GTC for their continued support for HiPERCAM in what have been very difficult circumstances over the past year.

%%%%%%%%%%%%%%%%%%%%%%%%%%%%%%%%%%%%%%%%%%%%%%%%%%
\section*{Data Availability}

All data analysed in this work can be made available upon reasonable request to the authors.



\bibliographystyle{mnras}
\bibliography{j0240} % if your bibtex file is called example.bib

%\begin{thebibliography}{99}
%\end{thebibliography}

%%%%%%%%%%%%%%%%%%%%%%%%%%%%%%%%%%%%%%%%%%%%%%%%%%

%%%%%%%%%%%%%%%%% APPENDICES %%%%%%%%%%%%%%%%%%%%%

%\appendix



%%%%%%%%%%%%%%%%%%%%%%%%%%%%%%%%%%%%%%%%%%%%%%%%%%

% Don't change these lines
\bsp	% typesetting comment
\label{lastpage}
\end{document}